\documentclass[prb,12pt,onecolumn]{revtex4}
\usepackage{amsmath}
\usepackage{graphicx}
\usepackage{dcolumn}
\usepackage{bm}

\input{tcilatex}

\begin{document}

\title{Variational method study of the spin-1 Ising model with biaxial
crystal-field anisotropy}
\author{J. Ricardo de Sousa$^{1,2}$ and N. S. Branco$^{3}$}
\affiliation{$^{1}$Departamento de F\'{\i}sica, Universidade Federal do Amazonas\\
3000, Japiim, 69077-000, Manaus-AM, Brazil\\
$^{2}$Departamento de F\'{\i}sica, ICEx, Universidade Federal de Minas Gerais%
\\
Av. Ant\^{o}nio Carlos 6627, CP 702, 30123-970, Belo Horizonte-MG, Brazil\\
$^{3}$ Departamento de F\'{\i}sica, Universidade Federal de Santa Catarina,
88040-900, Florian\'{o}polis-SC, Brazil}

\begin{abstract}
The phase diagram of the spin-1 Ising model in the presence of a biaxial
crystal-field anisotropy is studied within the framework of a variational
approach, based on the Bogolyubov inequality for the free energy.
We have investigated the effects of a transverse
crystal-field $D_{y}$ on the phase diagram in the $T-D_{x}$ plane. 
Results obtained by using effective-field theory (EFT) on the honeycomb (%
$z=3$), square ($z=4$) and simple cubic ($z=6$) lattices ($z$ is the
coordination number) show only continuous phase transitions, while
the variational approach presents first-order and continuous phase transitions for $D_{y}=0$%
. We have also used the EFT for larger values of $z$ and we observe the
presence of tricritical points in the phase diagrams, for $z\geq 7$, in
accordance with the variational approach results.

Pacs number: 05.70.-a, 64.60.-i, 75.10.Nr; 75.50.Lk
\end{abstract}

\maketitle

In the past few decades there has been an increasing number of studies on
the phase transition of the spin-1 Ising model with a longitudinal
crystal-field\cite{1,2,3,4,5,6,7}, which is described by the following
Hamiltonian:

\begin{equation}
{\mathcal H}=-J \sum_{<i,j>}S_{i}^{z}S_{j}^{z} + D 
\sum_{i}(S_{i}^{z})^{2},  \label{eq:hamilBC}
\end{equation}%
where the first sum is over pairs of nearest-neighbor spins, the second
sum is over all spins on the lattice, $J$ is the
exchange interaction, $D$ represents the longitudinal crystal-field, and $%
S_{i}^{z}$ is the $z$-component of a spin-$1$ operator at site $i$. The previous
Hamiltonian defines the so-called Blume-Capel (BC) model\cite{8} and has been studied
by various methods\cite{1,2,3,4,5,6,7,8}.

It is well known that when the uniaxial anisotropy, $D$, is greater than
$D_{c}=zJ/2$ ($z$ is the coordination number), the system is always in the
paramagnetic (\textbf{P}) phase, no matter the temperature. For
$D < D_{c}$, there is a ferromagnetic (\textbf{F}) state at low enough
temperatures. At $D=D_{c}$, we have a first-order
phase transition between the \textbf{F} and \textbf{P} phases. For null
anisotropy ($D=0$), the phase diagram in the ($T-D$) plane,
in two and three-dimensional lattices, presents a
continuous phase transition. As the anisotropy parameter $D$ increases,
the critical temperature of this transition decreases. At low temperatures, the system presents
a first-order phase transition, with a tricritical point (\textbf{TCP})
separating these two regimes.

For infinite dimension ($d\rightarrow \infty $), the classical model, Eq. (\ref{eq:hamilBC}),
can be exactly solvable, replacing $J$ by $J/N$ ($N$ is the total number of sites)
and the first sum running now over all pairs. This model with long-range
interaction is equivalent to
using mean-field approximation (MFA) or Curie-Weiss theory. The phase
diagram obtained by MFA is qualitatively equivalent to the rigorous results
of Monte Carlo simulation\cite{7} in a three-dimensional lattice.
The limit $d\rightarrow \infty $ corresponds to the MFA solution and,
therefore, all different methods used to treat the BC model will tend to
the MFA results when the coordination number $z$ is increased.

On the other hand, some studies have dealt with the effect of a transverse
crystal-field\cite{9,10,11,12,13}. In particular, the presence of a
transverse crystal-field in the BC model may change the nature of
the phase transition, due to quantum effects. It is known that
the determination of thermodynamic
properties of quantum models is a non-trivial task, mainly due to the
non-commutativity of the operators in the Hamiltonian. Therefore, the
use of aproximate methods in these classes of models is always 
relevant. Moreover, mean-field-like procedures are known to be
a good first approximation in describing
critical phenomena in three-dimensional models. We
then apply a variational approach to study the Hamiltonian:
\begin{equation}
\mathcal{H}=-J\sum_{<i,j>}S_{i}^{z}S_{j}^{z} + D
\sum_{i}(S_{i}^{z})^{2} + D^{\prime }\sum_{i}(S_{i}^{x})^{2},
\label{eq:transverse1}
\end{equation}%
where $D^{\prime }$ is the transverse crystal-field anisotropy.
Using the spin identity $\left( S_{i}^{x}\right) ^{2}+\left(
S_{i}^{y}\right) ^{2}+\left( S_{i}^{z}\right) ^{2}=S(S+1)$, one can rewrite
Eq. (\ref{eq:transverse1}) as%
\begin{equation}
\mathcal{H}=-J\sum\limits_{<i,j>}S_{i}^{z}S_{j}^{z}-D_{x}\sum%
\limits_{i}(S_{i}^{x})^{2}-D_{y}\sum\limits_{i}(S_{i}^{y})^{2},  \label{eq:transverse2}
\end{equation}%
where $D_{x}=D-D^{\prime }$, $D_{y}=D$ and an irrelevant constant term
was dropped out. In particular, for $D^{\prime
}=0$ or $D_{x}=D_{y}$ the quantum Hamiltonian, Eq. (\ref{eq:transverse2}), reduces to the purely
classical model, Eq. (\ref{eq:hamilBC}).

From the experimental point of view, the anisotropy defined in Eq. 
(\ref{eq:transverse2}) was shown to play an important role when the
thickness of the nonmagnetic TaN interlayer in FeTaN/TaN/FeTaN sandwiches
\cite{14} is changed.


On the other hand, theoretical studies of the Hamiltonian 
described by Eq. (\ref{eq:transverse2}), 
using effective-field theory 
(EFT) \cite{9,10,11,12,13} on
honeycomb ($z=3$), square ($z=4$) and simple cubic ($z=6$) lattices, have
shown that the system presents only continuous transitions, such that the
critical temperature approaches zero at two symmetrical transverse
crystal-fields, $D_{1x}=-D_{2x}$, when $D_{y}\leq 0$. It was observed that,
increasing $D_{x}$ from negative values, $T_{c}$ increases from $%
T_{c}=0$ at $D_{x}=D_{1x}$, passes through a maximum at $D_{x}<0$ and
vanishes again at a positive value of $D_{x}=D_{2x}$. Another result
observed in studies of Hamiltonian (\ref{eq:transverse2}) is the presence of a TCP for $D_{y}>0$. An
important and interesting question is to know if, for $D_{y}=0$, model (\ref{eq:transverse2})
has a TCP in the phase diagram in the $T-D_{x}$ plane, for large values of $z$.

So, we study here the model described by Eq.(\ref{eq:transverse2}), which we treat by
employing the variational approach based on the Bogolyubov inequality for the
free energy. The Bogolyubov variational principle for the free energy is
given by 
\begin{equation}
\mathcal{F}\leq \mathcal{F}_{o}+\left\langle \mathcal{H-H}_{o}(\eta
)\right\rangle _{o}\equiv \Phi (\eta ),  \label{eq:bogoliubov}
\end{equation}%
where $\mathcal{H}$ is the Hamiltonian of the model we want to treat, $%
\mathcal{H}_{o}$ is the trial Hamiltonian, which can be exactly solved and depends
on the variational parameter $\eta $, $\mathcal{F}_{o}$ is the free energy
associated with $\mathcal{H}_{o}$, and $\left\langle \cdot \cdot \cdot
\right\rangle _{o}$ is the thermal average over the ensemble defined by $%
\mathcal{H}_{o}$. The approximate free energy is given by the minimum of $%
\Phi (\eta )$ with respect to $\eta$.

We use a trial Hamiltonian which includes clusters of one spin, namely:
\begin{equation}
\mathcal{H}_{o}(\eta )=-\eta
\sum\limits_{i}S_{i}^{z}-D_{x}\sum\limits_{i}(S_{i}^{x})^{2}-D_{y}\sum%
\limits_{i}(S_{i}^{y})^{2},  \label{eq:sitiounico}
\end{equation}

Using Eqs. (\ref{eq:transverse2}) and (\ref{eq:sitiounico}) into Eq. 
(\ref{eq:bogoliubov}), we obtain the variational free
energy per spin:
\begin{equation}
\overline{\Phi }(\eta )\equiv \frac{\Phi (\eta )}{N}=\eta m-zJ\frac{m^{2}}{2}%
-k_{B}T\ln Z_{o}(\eta ),  \label{eq:energialivre}
\end{equation}%
with the partition function%
\begin{eqnarray}
Z_{o} &=&Tr\left\{ e^{\beta \left[ \eta
S_{i}^{z}+D_{x}(S_{i}^{x})^{2}+D_{y}(S_{i}^{y})^{2}\right] }\right\}
\label{7} \\
&=&\exp (2\beta D)+2\exp (\beta D)\cosh \left( \beta \sqrt{\Delta ^{2}+\eta
^{2}}\right) ,  \notag
\end{eqnarray}%
where $D=(D_{x}+D_{y})/2$, $\Delta =(D_{x}-D_{y})/2$, and $m=\left\langle 
\frac{1}{N}\sum\limits_{i}S_{i}^{z}\right\rangle _{o}$ is the
magnetization per spin.

Minimizing Eq. (\ref{eq:energialivre}), we obtain the variational parameter $\eta =zJm$, and
the Landau free energy is given by:
\begin{eqnarray}
\Psi (m) \equiv &-\beta \overline{\Phi }(\eta )  \label{eq:energialivredelandau} \\
&=&-zK\frac{m^{2}}{2}+\ln \left[ \exp (2\beta D)+2\exp (\beta D)\cosh \left(
W\right) \right] ,  \notag
\end{eqnarray}%
with the equation of state (magnetization):
\begin{equation}
m=\left( \frac{zKm}{W}\right) \frac{2\sinh \left( W\right) }{\exp (\beta
D)+2\cosh \left( W\right) },  \label{eq:magnetizacao}
\end{equation}%
where $K=\beta J$ and $W=\sqrt{(\beta \Delta )^{2}+(zKm)^{2}}$.

To analyze the second-order phase transition and the 
TCP, we expand the free energy in powers of $m$:
\begin{equation}
\Psi (m)\simeq a_{2}m^{2}+a_{4}m^{4}+a_{6}m^{6}+\cdot \cdot \cdot  \label{eq:expansao}
\end{equation}%
where the coefficients $a_{p}=\frac{1}{p!}\left( \frac{\partial ^{p}\Psi }{%
\partial m^{p}}\right) _{m=0}$ are functions of the parameters $D_{x}$, $%
D_{y}$ and $K$.

The continuous phase transition occurs with the breakdown of the order
parameter $m$ (i. e., $m\rightarrow 0$). So the phase transition boundaries
of the \textbf{F} phase are determined from the zero point of the
coefficient of the second-order term in Eq. (\ref{eq:expansao}), i.e., $a_{2}=0$ and $%
a_{4}>0$. To obtain the phase diagram using MFA in all interval of the
parameters $D_{x}$ and $D_{y}$, first we compute the free energies, Eq.
(\ref{eq:energialivredelandau}), for the \textbf{F} ($m\neq 0$) and \textbf{P} ($m=0$) phases. All the results
presented in this section were obtained from numerical solutions of Eqs. (\ref{eq:energialivredelandau}) and
(\ref{eq:magnetizacao}). First-order transitions correspond then to the locus on the
phase diagram where free energies are equal (i.e., $\Psi _{F}(m)=\Psi
_{P}(0) $), with $m=m_{c}\neq 0$, while for continuous phase transitions,
$m_{c}=0$. The TCP - in
which the phase transition changes from continuous to first-order -
is determined by $a_{2}=a_{4}=0$ and $a_{6}>0$.

Defining the dimensionless parameters $\tau \equiv k_{B}T/zJ$, $\delta
_{x}=D_{x}/zJ$, and $\delta _{y}=D_{y}/zJ$, the numerical results for the
phase diagram in the $\tau -\delta _{x}$ plane, for different values of $%
\delta _{y}$, are shown in Fig. 1. We chose to depict the values $\delta _{y}=0.5$ 
(curve (a)), $0.0$ (curve (b)) and $-0.50$ (curve (c)). In this
figure, the solid and traced lines correspond to continuous and first-order
phase transitions respectively. The filled dots are the TCP. Within the
studied region of transverse crystal-field $\delta _{x}$, we could not find
any point satisfying the TCP condition ($a_{2}=a_{4}=0$) for the model with 
$\delta _{y}<0$. The situation is different for $\delta _{y}\geq 0$:
when we increase $\delta _{x}$ from negative values
the transition temperature $\tau _{c}$ increases, passes through a maximum and
satisfies the TCP condition at the points ($\delta _{xT},\tau _{T}$)=($%
0.9472,0.2845$) and ($0.4243,0.3330$) for $\delta _{y}=0$ and $0.50$,
respectively. Thus, we may conclude that the Ising model with a transverse
crystal-field exhibits a TCP on the $\tau $ versus $\delta _{x}$ curve, when
the transverse crystal field $\delta _{y}$ is null. This is in contrast to
EFT results for coordination numbers $z=3$, $4$ and $6$ \cite{9,10,11,12,13}, 
for which no TCP is found for any value of $\delta_{y}$.
.

\FRAME{ftbphFU}{5.2486in}{4.4079in}{0pt}{\Qcb{Phase diagram in the $\protect%
\tau -\protect\delta _{x}$ plane for the spin-1 Ising model with a biaxial
crystal-field using MFA, for the following values of $\protect\delta _{y}$:
$0.50$ (a), $0.0$ (b) and $-0.5$ (c). The continuous and
traced lines correspond to continuous and first-order phase transitions,
respectively. Filled dots represent tricritical points. The inset shows a
comparison of the first-order line obtained by using the free energy (traced
line) and condition $a_{2}=0$ and $a_{4}<0$, which corresponds to an
unstable solution for $\protect\delta _{y}=0$ (dotted line).}}{}{Figure}{%
\special{language "Scientific Word";type "GRAPHIC";maintain-aspect-ratio
TRUE;display "USEDEF";valid_file "T";width 5.2486in;height 4.4079in;depth
0pt;original-width 4.0153in;original-height 3.3702in;cropleft "0";croptop
"1";cropright "1";cropbottom "0";tempfilename
'JNX7H508.wmf';tempfile-properties "XPR";}}

Some authors\cite{12,13} have used the condition $a_{2}=0$ and $a_{4}<0$ to
determinate the first-order line, but this condition corresponds to the 
stability line of the \textbf{P} phase. One can note that it is not possible to
calculate first-order lines based on the equation of state (\ref{eq:magnetizacao}) alone,
because in this case one has $m\neq 0$ at the transition point. Therefore,
the first-order line is not correctly obtained using the
EFT \cite{9,10,11,12,13}: only critical lines and TCP can be
obtained with this procedure, since an expression for the free energy is not available.  
In order to illustrate the
difference between the two procedures, when it comes to
locate first-order lines, we show in the inset of Fig. 1 the
phase diagram for $\delta _{y}=0$, using the free energy method (traced
line, \textbf{method I}) and the condition $a_{2}=0$ and $a_{4}<0$ (dotted
line, \textbf{method II}). We note, for example, that at zero temperature ($%
\tau =0$), $\delta _{xc}=1.0$ and $\delta _{xc}=2.0$ from \textbf{%
methods I} and \textbf{II}, respectively.

To access the influence of quantum effects, we
can compare our results for $\delta _{y}=0$ with the value of the TCP 
for the Blume-Capel model using
MFA, which is equivalent to replacing $\delta _{x}=\delta _{y}=\delta $ in
Eqs. (\ref{eq:magnetizacao}) and (\ref{eq:expansao}); for the latter, the coordinates 
of the TCP are ($\delta
_{T}=0.462$, $\tau _{T}=0.333$)\cite{8}. This result for the TCP in the
Blume-Capel model is different from the one on the Ising model with a
transverse crystal-field and $\delta _{y}=0$, namely ($\delta
_{xT}=0.9472,\tau _{T}=0.2845$). In order to illustrate the difference
between the two models, in Fig. 2 we show the phase diagram in the $\tau
-\delta $ plane for the longitudinal (curve (b), where $\delta
_{x}=\delta _{y}=\delta $) and transverse (curve (a), where $\delta
_{y}=0$ and $\delta _{x}=\delta $) crystal-field cases. 
Particularly, the result for the transition temperature at 
$\delta =0$ is $\tau _{c}=2/3$, and for $\tau =0$ (ground state),
we obtain $\delta _{c}=1/2$ and $2$ for the transverse  (a) and longitudinal
(b) Blume-Capel models, respectively. For negative values of
$\delta $, the transition line for the Blume-Capel model extends
to $\delta \rightarrow -\infty $ and its critical temperature is always finite. The
system is in the ordered (ferromagnetic) phase for $\tau <\tau
_{c}(\delta)$ while it
is in a disordered (paramagnetic) phase for $\tau >\tau _{c}(\delta)$. On the
other hand, when the transverse crystal-field anisotropy $%
\delta \rightarrow -\infty $, we have a paramagnetic state in all
temperatures. In the case of a transverse crystal-field, the critical
temperature goes to zero at the quantum critical point $\delta _{c}=-2.0$,
which is symmetrical to the first-order phase transition point at positive $\delta$,
$\tau =0$, $%
\delta _{c}^{\prime }=2.0$. 


\FRAME{ftbphFU}{5.3264in}{4.3967in}{0pt}{\Qcb{Phase diagram in the $\protect%
\tau -\protect\delta $ plane for the spin-$1$ Ising model with longitudinal
(a) and transverse (b) crystal-field using MFA. The continuous and traced lines
correspond to continuous and first-order phase transitions, respectively.
The black points represents the TCP. \ }}{}{Figure}{\special{language
"Scientific Word";type "GRAPHIC";maintain-aspect-ratio TRUE;display
"USEDEF";valid_file "T";width 5.3264in;height 4.3967in;depth
0pt;original-width 4.0759in;original-height 3.3607in;cropleft "0";croptop
"1";cropright "1";cropbottom "0";tempfilename
'JNX7AA07.wmf';tempfile-properties "XPR";}}

We have extended the EFT approach on model (\ref{eq:transverse2}), to investigate the possible
existence of a TCP for some values of the
coordination number. We note the absence of this point
in the region $\delta _{y}\leq 0$ and for values of $z\leq 6$,
in accordance with previous results \cite{9,10,11,12,13}. However,
we verify the presence of a TCP for lattices with $z\geq 7$. In order to
illustrate the presence of this TCP, we analyze the bcc ($z=8$) and fcc ($%
z=12$) lattices and, by using the condition $a_{2}=a_{4}=0$ and $a_{6}>0$, we
obtain the coordinates of the TCP: ($\delta _{xT},\tau _{T}$)= ($%
1.0817,0.1708 $) and ($1.0111,0.2177$), respectively. We compare the results
of the TCP obtained by EFT with the MFA values ($0.9472,0.2845$) and we
verify that, with the increase of the coordination number, the TCP tends
to the MFA values, as expected.

In summary, we present mean-field calculations based on the Bogolyubov
inequality for the free energy, to obtain the phase diagram of the spin-$1$ Ising
model with biaxial crystal-field anisotropy. We verify that for zero transverse
crystal-field $\delta _{y}$, the system exhibit a TCP in the phase
diagram for all values of $z$, while the results obtained by effective
field theory\cite{9,10,11,12,13} for coordination number $z=3,4$ and $6$ present
only continuous phase transitions. We generalize the EFT
for lattices with large values of $z$, and it was observed the presence of a TCP for $%
z\geq 7$. With the increase of $z$, the EFT results tends to the MFA
solution. In order to investigate the MFA results by using the
differential operator technique, we suggest the use of the correlated-EFT\cite%
{17} or the EFT in a finite cluster with two spins\cite{18}. Finally, owing to the
mathematical simplicity of our formulation, we hope that the present work
will be potentially useful for studying more complicated systems, including, for
example, a transverse magnetic field. These generalizations are
under investigation.

\textbf{ACKNOWLEDGMENTS: }The authors acknowledge valuable discussions with
Dr. J. A. Plascak. This
work was partially supported by CNPq, FAPEAM and FAPESC (Brazilian agencies).

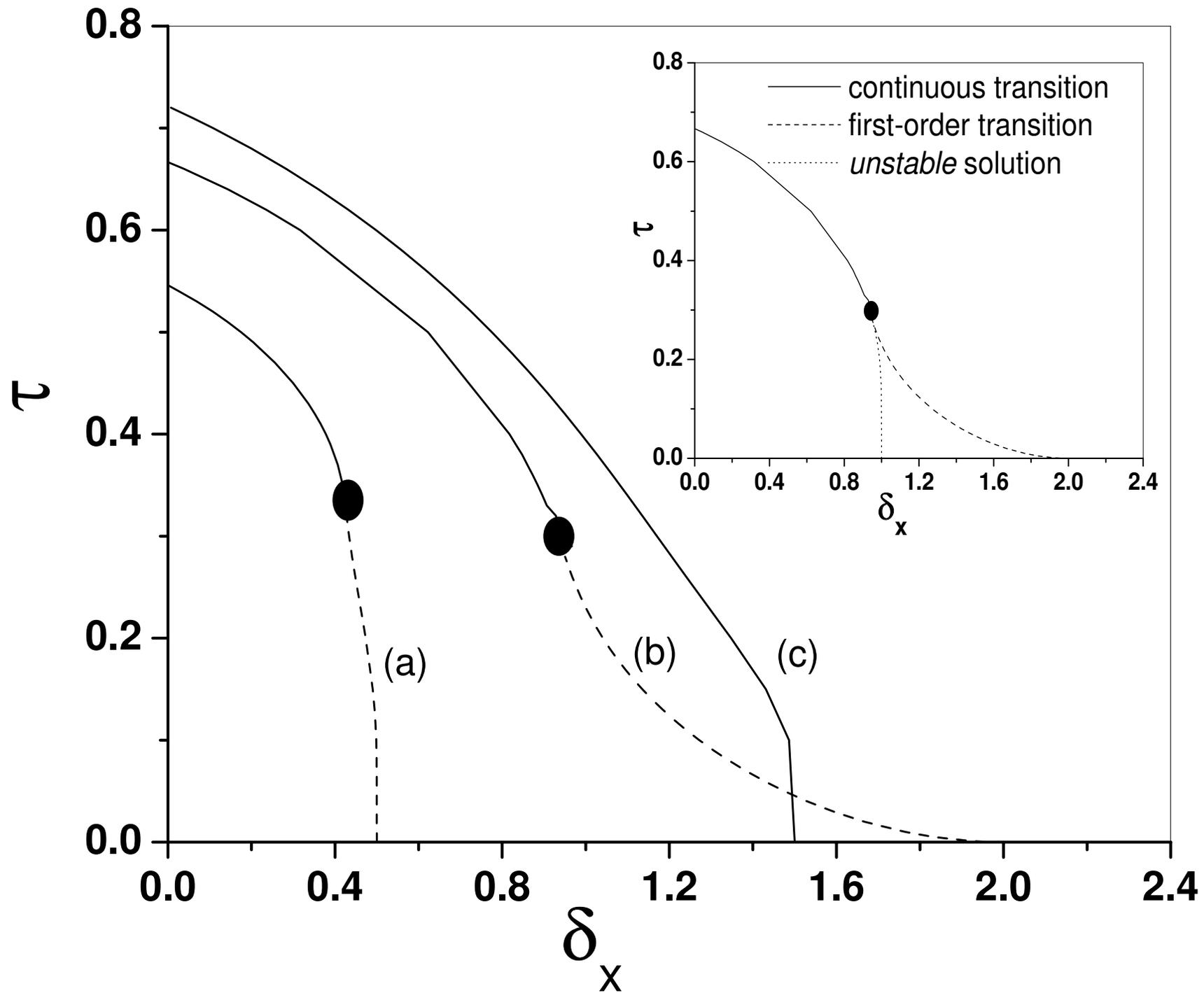
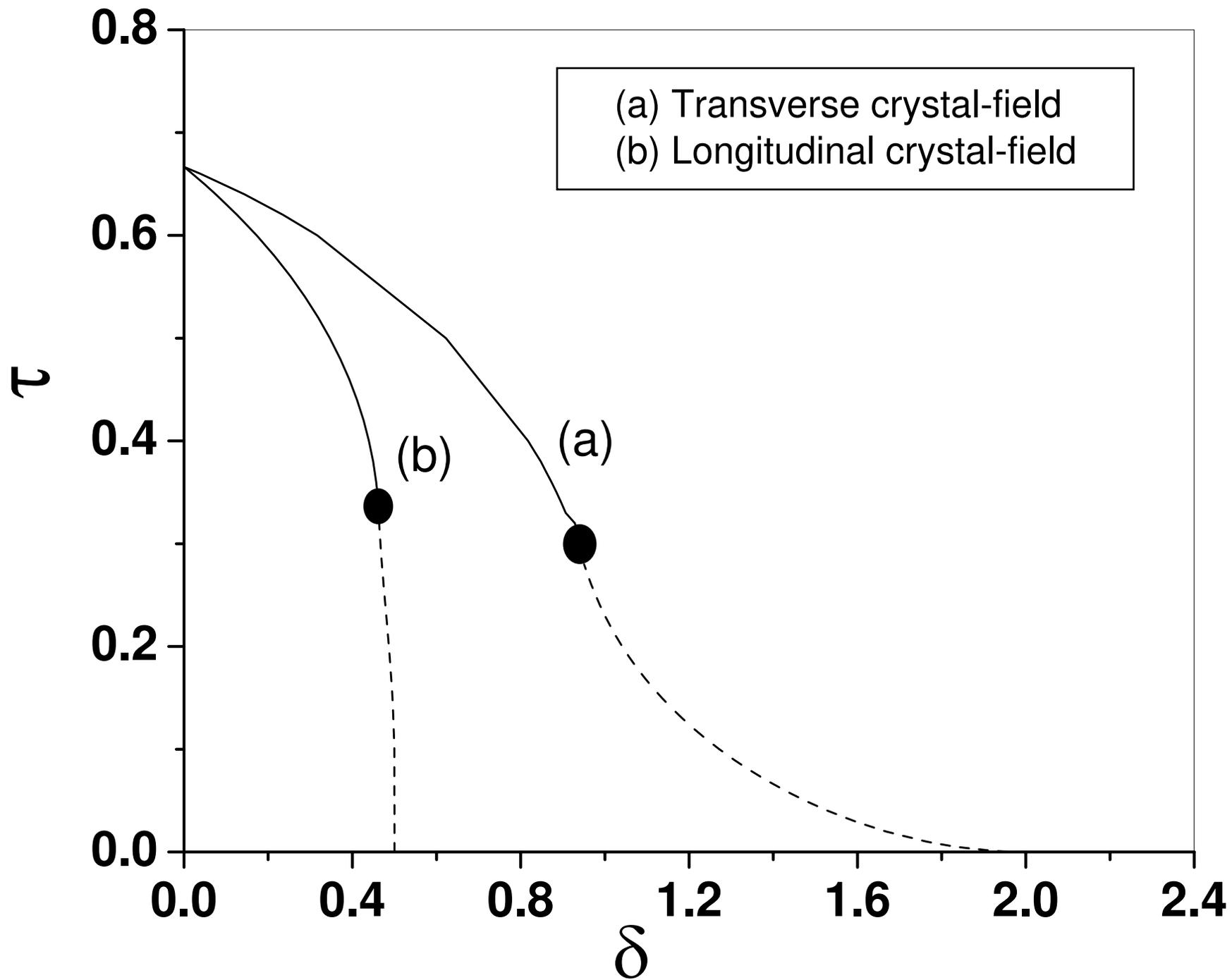
\end{document}